\documentclass[apl,prl,twocolumn,groupedaddress]{revtex4}
\usepackage{graphicx}
\usepackage{calc}
\usepackage{pifont}
\usepackage{amssymb}
\usepackage{epsfig}
\usepackage{amssymb}
\usepackage{amsmath}
\usepackage{color}
\DeclareMathAlphabet{\mathitb}{OT1}{cmr}{bx}{sl}
\begin{document}

\renewcommand{\thefootnote}{\fnsymbol{footnote}}
\title{Sputtered Gold as an Effective Schottky Gate for Strained Si/SiGe Nanostructures}
\author{G. D. Scott$^1$}
\email{gscott@physics.ucla.edu}
\author{M. Xiao$^1$}
\author{E. T. Croke$^2$}
\author{E. Yablonovitch$^3$}
\author{H. W. Jiang$^1$}

\affiliation{
$^1$Department of Physics and Astronomy, UCLA, Los Angeles, CA 90095\\
$^2$HRL Laboratories, LLC, Malibu, CA 90265\\
$^3$Department of Electrical Engineering, UCLA, Los Angeles, CA
90095}

\date{\today}

\begin{abstract}

Metallization of Schottky surface gates by sputtering Au on strained
Si/SiGe heterojunctions enables the depletion of the two dimensional
electron gas (2DEG) at a relatively small voltage while maintaining
an extremely low level of leakage current.  A fabrication process
has been developed to enable the formation of sub-micron Au
electrodes sputtered onto Si/SiGe without the need of a wetting
layer.

\end{abstract}

\maketitle

Significant effort has been directed toward the development of
electrostatically defined quantum dots as potential elements for
quantum computation information.  While a high level of control and
sophistication has been achieved in current GaAs/AlGaAs based
structures,\cite{GaAs} silicon based heterostructures are expected
to have the distinct advantage of possessing extremely long electron
phase coherence lifetimes,\cite{Vjijen2000,Tyryshkin2003} which can
be attributed to the small spin-orbit interaction and the low
natural abundance of isotopes with nuclear spin.

Means of control in lateral quantum dot devices is often exercised
through the use of Schottky barrier top gates in which metal
electrodes patterned on the semiconductor surface capacitively
couple to the 2DEG. By applying a bias on the gates one can
selectively deplete the charge carriers in the 2DEG directly below,
and in the vicinity of, the gates, thereby controlling current flow.
Over the last several years attempts were made to create mesoscopic
devices on strained Si/SiGe heterostructures by directly mimicking
the existing geometries and fabrication processes that have been
employed on GaAs/AlGaAs based heterostructures. The success was
often limited due to the high level of leakage current and/or the
incomplete depletion of the 2DEG by Schottky gates on strained
Si/SiGe. Several innovative approaches have recently been introduced
as means of circumventing these
obstacles.\cite{Bo2002,Klein2004,Sakr2005} These methods primarily
focus on the formation of side gates through various etching
techniques.  Unfortunately, in order to reach a comparable level of
stability, control, and effectiveness of confinement achieved in
GaAs based structures,  devices using multiple surface gates will
most likely be required.  Towards this end, both Berer et
al.\cite{Berer2005} and Slinker et al.\cite{Slinker2005} have used
evaporated Pd on strained Si/SiGe as Schottky gates with positive
results. Despite this encouraging development, it is recognized that
further improvement of the effectiveness of the Schottky gates is
needed to gain control in the few-electron regime.

In this paper, we demonstrate that significant improvements to
Schottky gates on Si/SiGe can be made by the proper choice of gate
metal and a simple change in the mechanism of gate metal deposition.
In particular, we found that the Schottky gates formed by sputtering
Au on strained Si/SiGe heterojunctions enables the depletion of the
two dimensional electron gas (2DEG) at a relatively small voltage
while maintaining an extremely low level of leakage current.

The devices used for our experiment were fabricated on modulation
doped Si/SiGe heterostructures, which are commonly used for the
fabrication of lateral quantum dots.  These samples have a peak
mobility of about $10^5 cm^2/{V-sec}$ at a density of
$3x10^{11}/cm^2$, and are grown by MBE with a layer by layer
composition as follows: Si(5nm)/SiGe$_{0.25}$(35nm)/$\delta$-dopedSb
$=2x10^{12}cm^{-2}$/SiGe$_{0.25}$(22.5nm)/Si(15nm)/SiGe$_{0.25}$(1$\mu$m)/
SiGe$_{0.20-0.25}$(0.9$\mu$m)/SiGe$_{0.108-0.20}$(1.5$\mu$m)/
SiGe$_{0.018-0.108}$(1.5$\mu$m)/Si(100nm)/ p-Si(100) substrate.

To test the effectiveness of various Schottky gates, a large square
gate ($150{\mu}m \times 150{\mu}m $) was defined by UV lithography
atop a chemically etched mesa.  Ohmic contacts to the 2DEG were
formed by phosphors ion implantation.  Pd, Cr and Au films were
deposited as gate metals by evaporation.  The coupling strength
between the gate and 2DEG (i.e. depletion voltage) as well as
leakage between the gate and 2DEG were both measured as a function
of applied voltage (Fig 1). The results of the leakage current from
the evaporated gates are in line with the known Schottky barrier
values. For Pd the onset of measurable leakage began at a larger
value of applied bias, in accordance with the larger Schottky
barrier it forms with n-type silicon. However, in terms of
effectiveness of gate-coupling, Pd required a correspondingly larger
value of bias to fully deplete the 2DEG. The strongest coupling
between gate and 2DEG was found with Au, which required only a very
small voltage for the complete depletion of the 2DEG, but had
unacceptably high levels of leakage current.

\begin{figure*}[!t]
\begin{center}
\includegraphics [scale = 0.25]{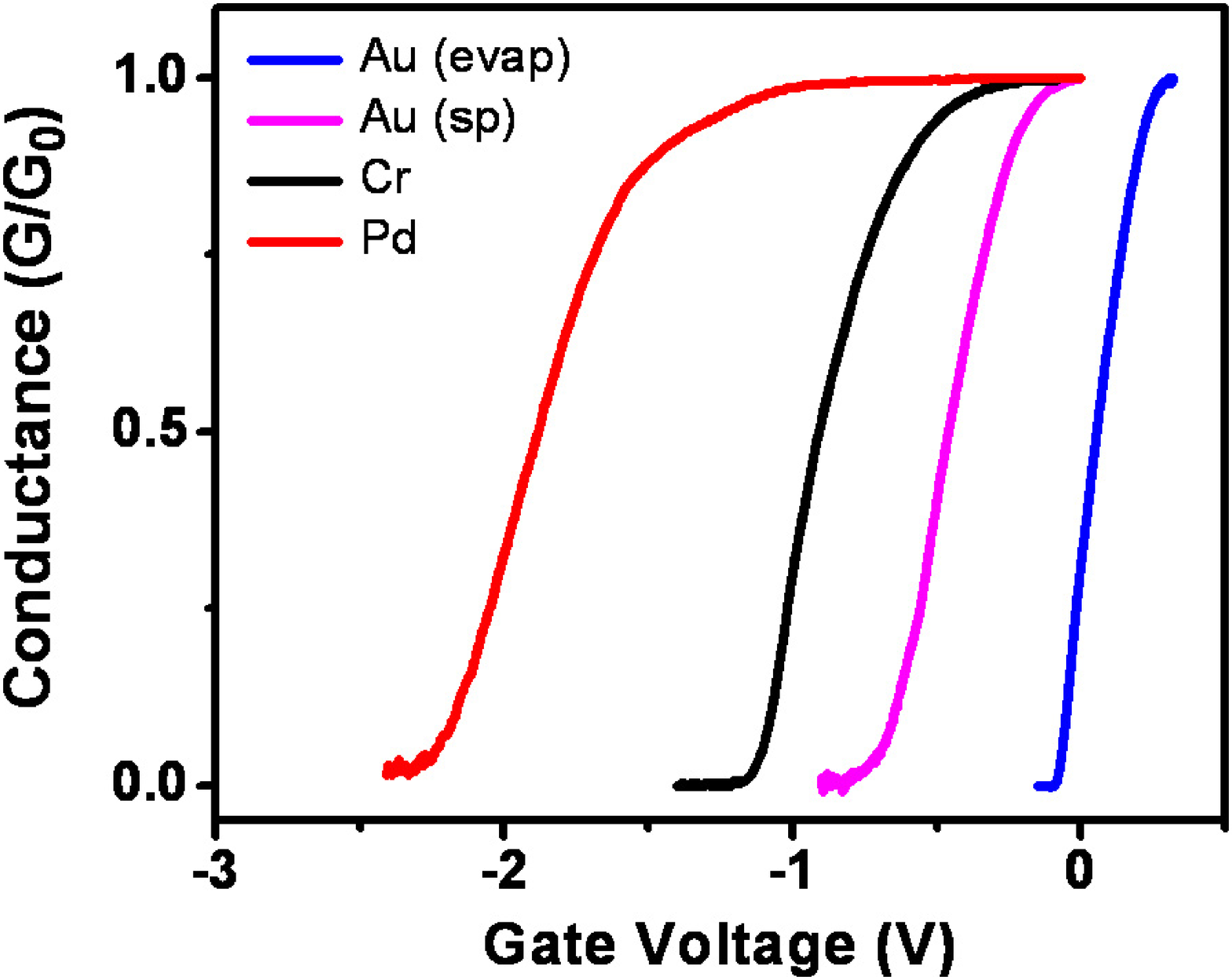}
\hspace {7mm} \includegraphics [scale = 0.25]{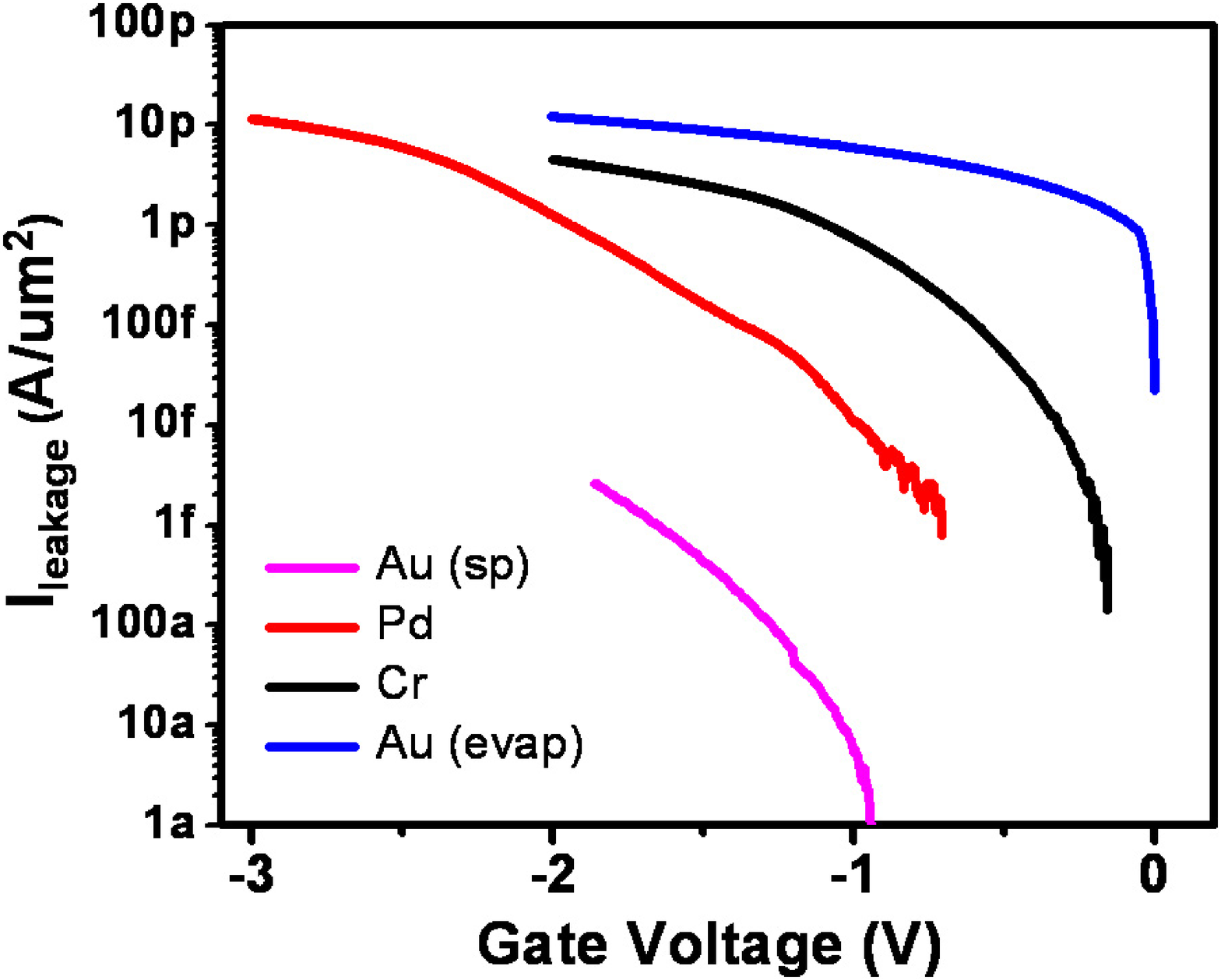}
\end{center}
\hspace\fill \vspace{-10mm} \caption{a) Conductance measured as a
function of applied gate voltage for four different gate
compositions.  b) Leakage current per square micron of gate
electrode measured as a function of applied voltage for the same
four samples. Gates of sputtered Au demonstrate levels of leakage
several orders of magnitude lower, on average, and possess strong
coupling to the 2DEG.} \label{Graphs2}
\end{figure*}

The Schottky-Mott theory predicts that the Schottky barrier height,
${\Phi}_B$, will depend linearly on the metal work function
${\Phi}_B = ({\Phi}_m - {\chi}_s)$, where ${\Phi}_m$ is the work
function of the metal and ${\chi}_s$ is the electron affinity of the
semiconductor.  This relation does not always hold in practice. When
a metal and a semiconductor are joined together, ${\Phi}_B$ will
depend sensitively not only on the identity of the materials, but
also on the details of the interface structure that results. In
covalently bonded semiconductors, such as silicon, surface gates may
be unable to effectively modulate the density of the underlying 2DEG
due to localized surface charges. The large values of applied
voltage necessary to cause depletion with Pd and Cr gates are
indicative of correspondingly large surface charge densities.

The choice of gate metal will influence the relative magnitude of
${\Phi}_B$, and a larger ${\Phi}_B$ dictates a lower level of
leakage current from the metal to the semiconductor at a particular
bias. But even though Pd, for example, has a relatively large
${\Phi}_B$, a large voltage is also required to achieve depletion.
Higher operating biases lead to increased levels of leakage current,
which in turn compromise the surface gate characteristics.

We also deposited gold by sputtering instead of evaporation. Gates
with a sputtered layer of gold always demonstrated dramatically less
leakage current as compared to the evaporated gates, while
maintaining excellent coupling to the 2DEG (Fig 1). The kinetic
energy of evaporated atoms corresponds to the temperature of the
evaporating surface, and is therefore typically around 0.1 - 0.3 ev.
The energy of sputtered atoms is roughly 100 times greater (for Au:
$\sim$ 8-16 ev).\cite{Wehner1959,Turner1992,Satpati2005} According
to the software program ``Stopping and Range of Ions in Matter''
(SRIM), sputtered gold atoms may penetrate into silicon to a depth
of around 1 nm, while evaporated atoms will not penetrate at all.
The ability of sputtered gold atoms to penetrate the silicon cap
will promote an interdiffusion of gold and silicon, possibly
triggering the formation of gold silicide
compounds.\cite{Zavodinsky1997,Sharma1984}  We speculate that, much
akin to platinum silicide,\cite{Streetman1980} gold silicide may
form a larger, more stable Schottky barrier with n-type silicon when
compared to bare gold.

Unfortunately, thin films of Au will not customarily adhere well to
silicon due to the SiO$_2$ that forms on the surface, causing
serious problems during the liftoff phase of the fabrication
procedure.  By removing the native oxide layer with a buffered oxide
etch (NH$_3$F:HF, 6:1), we found it possible to attain sufficient
adhesion enabling us to repeatedly and reproducibly form devices out
of evaporated and sputtered Au without any liftoff problems.

\begin{figure}[!h]
\begin{center}
\includegraphics [scale = 0.32]{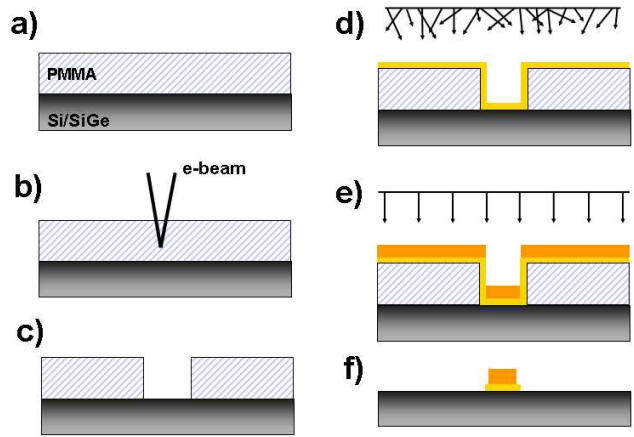}
\end{center}
\hspace\fill \vspace{-5mm} \caption{Fabrication Steps.  a) The
substrate is spin-coated with 495 PMMA.  b) Electron beam
lithography is used to define the device pattern.  c) PMMA exposed
during the Ebeam writing process is removed after developing.  The
sample is then placed in a BOE solution for 15 seconds to remove
SiO$_2$ layer.  d) A thin layer of gold is sputtered onto the
surface.  To allow for liftoff, the sputtered layer must be kept to
a minimum thickness. e) A thicker layer of gold is evaporated on top
of the sputtered gold layer.  f) Liftoff in hot acetone leaves an
intact device with excellent leakage properties.}
\label{SputterFab2}
\end{figure}

Unlike the deposition of metal by evaporation, during the sputtering
process metal is not deposited from one direction (Fig 2d).  As
Argon ions bombard the sputtering target, gold atoms are deposited
from many different directions. The result is that sputtering may
generate complete coverage of the developed lithography pattern,
making liftoff extremely difficult, if not impossible.  This is
especially problematic for thinner resists and finer lithographic
structures, namely for electron beam lithography using PMMA resist.
Since the majority of nanoscale devices necessitate the use of
electron beam lithography, a fabrication technique was required that
would facilitate metallization by sputtering.

We found that if the sputtered layer is thin enough ($\lesssim$ 15
nm) it can still be lifted off cleanly.  The final devices were
produced by using a BOE to remove the SiO$_2$ immediately prior to
metallization.  We then sputter a thin layer (12 nm) of gold onto
the surface.  After sputtering, a thicker layer of gold (35 nm) is
deposited by electron beam evaporation to reinforce the structural
integrity of the device. Lift off was then performed in heated
acetone (55 C) for approximately 2 hours (Fig 2).

\begin{figure}[!h]
\begin{center}
\includegraphics [scale = 0.25]{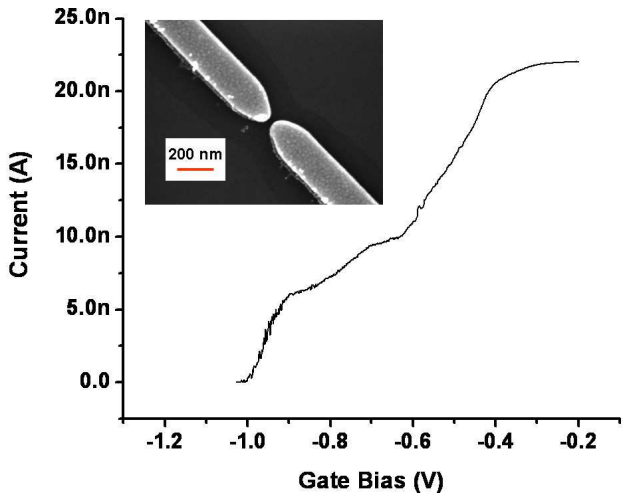}
\end{center}
\hspace\fill \vspace{-8mm} \caption{Current through a split gate,
fabricated with a sputtered layer of gold, as a function of applied
voltage.  Stepwise depopulation of edge channels as evinced through
decreasing steps in current, as well as a pinch off voltage around
-1.0 V indicate excellent gate characteristics. Inset: SEM picture
of the split gate structure.} \label{DepletionCurve}
\end{figure}

Complete pinch off of the current through split gate structures in
devices formed by this fabrication technique (Fig 3 inset) was
observed to occur at biases in accordance with Fig 1a, indicating a
strong coupling between the gate electrodes and the 2DEG. The
emergence of more pronounced steps in the depletion curves,
demonstrating the stepwise depopulation of available edge channels,
is evidence of sharp electrostatic confinement beneath the gates.
Both the lower operating bias necessary to achieve depletion, and
the enormous reduction of leakage current lead to a more stable
electrostatic environment, which will be necessary for developing
the next generation of SiGe based devices.

By using Schottky surface gates formed with sputtered Au, we
demonstrated that it is possible to dramatically reduce the leakage
current in strained Si/SiGe while simultaneously improving the
coupling strength between the gates and 2DEG. The advantages of
sputtering, as opposed to thermal or electron beam evaporation, are
likely due to the interdiffusion of Au and Si leading to the
formation of gold-silicide compounds. A nano-fabrication process is
developed specifically to enable the fabrication of Au sputtered
sub-micron gates. A split gate device using this technique
demonstrates excellent electrostatic characteristics. Advancements
in the effectiveness of Schottky gates allows for the continued use
of conventional surface gate geometries, superseding the need for
more exotic gating techniques.

The authors would like to thank Thomas Szkopek for many insightful
discussions.  This work is supported by the Defense MicroElectronics
Activity (DMEA 90-02-2-0217) and by MARCO MSD Center.

\end{document}